\documentclass[aps,pra,showpacs,preprintnumbers,superscriptaddress]{revtex4}
\usepackage{amsmath}
\usepackage{amssymb}
\usepackage{graphicx}
\usepackage{dcolumn}
\usepackage{bm}

\input{tcilatex}

\begin{document}

\title{Linear Optical Implemention of a Quantum Network for Quantum
Estimation}
\author{Zhi-Wei Wang }
\affiliation{Key Laboratory of Quantum Information, University of Science and Technology
of China, CAS, Hefei 230026, People's Republic of China}
\author{Jian Li}
\email{jianli@seu.edu.cn}
\affiliation{Department of Physics, Southease University, Nanjing 211189, People's
Republic of China}
\affiliation{Key Laboratory of Quantum Information, University of Science and Technology
of China, CAS, Hefei 230026, People's Republic of China}
\author{Yun-Feng Huang}
\email{hyf@ustc.edu.cn}
\affiliation{Key Laboratory of Quantum Information, University of Science and Technology
of China, CAS, Hefei 230026, People's Republic of China}
\author{Yong-Sheng Zhang}
\affiliation{Key Laboratory of Quantum Information, University of Science and Technology
of China, CAS, Hefei 230026, People's Republic of China}
\author{Xi-Feng Ren}
\affiliation{Key Laboratory of Quantum Information, University of Science and Technology
of China, CAS, Hefei 230026, People's Republic of China}
\author{Pei Zhang}
\affiliation{Key Laboratory of Quantum Information, University of Science and Technology
of China, CAS, Hefei 230026, People's Republic of China}
\author{Guang-Can Guo}
\affiliation{Key Laboratory of Quantum Information, University of Science and Technology
of China, CAS, Hefei 230026, People's Republic of China}

\begin{abstract}
We present a scheme for simulating the quantum network of quantum estimation
proposed by A. K. Ekert \textit{et al}. [Phys. Rev. Lett. 88, 217901
(2002)]. We experimentally implement the scheme with linear optical
elements. We perform overlap measurements of two single-qubit states and
entanglement-witness measurements of some two-qubit states. In addition, it
can also be used for entanglement quantification for some kinds of states.
From the other perspective, we physically realize the positive but not
completely positive map, transposition.
\end{abstract}

\maketitle

A density matrix can completely characterize the state of a quantum system.
For low dimensional systems, quantum estimations of\ density matrices are
well mastered. Then the properties of the quantum state, which can be
quantified in terms of linear or nonlinear functionals of density matrices,
can be extracted from it. However, for high dimensional systems, the state
reconstruction becomes very difficult and a direct estimation of a specific
quantity may be more efficient. Thus one quantum network (shown in Fig. 1)
and its related idea are presented in \cite{E,1,F,4}, which can be used as a
basic building block for direct quantum estimations of both linear and
nonlinear functionals of any density operator. The network can perform
overlap measurements of two single-qubit states and entanglement-witness
measurements \cite{EW}\ of some two-qubit states. In addition, it can also
be used for entanglement quantification for some kinds of states.

In this paper, we put forward a scheme to simulate the network and
experimentally realize it using linear optical elements. In Fig. 1, the
network is a one-qubit interferometric setup (consisting of two Hadamard
gates and phase shift, followed by a measurement in the computational basis)
modified by inserting a controlled-SWAP operation between the two Hadamard
gates, with its control on the single qubit and SWAP on an unknown system
described by density operator $\rho _{a}\otimes \rho _{b}$. Our scheme is
based on two\ Hong-Ou-Mandel interferometers \cite{HOM}. The scheme set-up
is shown in Fig. 2 (a). A $\beta $-barium borate (BBO) crystal arranged in
the Kwiat type configuration \cite{kwiat} is pumped by a UV laser beam.
Through the spontaneous parametric down-conversion (SPDC) process, an
entangled state of the form

\begin{equation}
a|HH\rangle +b|VV\rangle \text{, }a\text{,}b\in R
\end{equation}%
can be produced \cite{kwiat}, where $H$ and $V$ represent horizontal and
vertical polarization of photons respectively. By means of this entanglement
source, some simple optical elements and optical set-ups, such as half wave
plate (HWP), quarter wave plate (QWP) and quartz, we can prepare some other
kinds of entangled states and mixed states \cite{mix}. Then we can put the
prepared states into the quantum network. The path of the two-photon is
selected as the control qubit while the target qubits have many choices,
such as polarization, spatial mode, and moment. For simplicity,\ here we
just encode the qubit with polarization of the photon.

In general, we suppose that after state preparation, the input states are $%
|\psi _{a}\rangle $ and $|\psi _{b}\rangle $. After the twin photon passes
through the two beam splitters (BS) BS$_{1}$, BS$_{2}$ and a phase shifter,
the twin photon state has the form (without normalization) 
\begin{equation}
|\psi _{a}\rangle (e^{i\varphi }|u_{1}\rangle -|d_{1}\rangle )|\psi
_{b}\rangle (|u_{2}\rangle +|d_{2}\rangle )\text{.}
\end{equation}%
Here the factor $e^{i\varphi }$ is due to the phase shifter and $u_{i}$ and $%
d_{i}$ ($i=1,2,3,4$) represent the two path of the photon after BS$_{i}$,
which is shown in Fig. 2. When they reach BS$_{\text{3}}$, BS$_{\text{4}}$,
we can obtain 
\begin{equation}
|\psi _{a}\rangle \left[ e^{i\varphi }\left( |u_{3}\rangle +|d_{3}\rangle
\right) -\left( |u_{4}\rangle +|d_{4}\rangle \right) \right] |\psi
_{b}\rangle \left[ \left( |u_{3}\rangle -|d_{3}\rangle \right) +\left(
|u_{4}\rangle -|d_{4}\rangle \right) \right] \text{.}
\end{equation}

Only considering the terms containing $u_{3}$ and $d_{4}$, we can obtain

\begin{equation*}
-e^{i\varphi}|\psi_{a}\rangle|u_{3}\rangle|\psi_{b}\rangle|d_{4}\rangle
-|\psi_{a}\rangle|d_{4}\rangle|\psi_{b}\rangle|u_{3}\rangle\text{.}
\end{equation*}
Then after coincidence in the sequence of$\ u_{3}$ and $d_{4}$, the final
state is 
\begin{equation}
|\psi^{\prime}\rangle=-e^{i\varphi}|\psi_{a}\rangle|\psi_{b}\rangle-|\psi
_{b}\rangle|\psi_{a}\rangle\text{.}
\end{equation}
In this way, we realize the control-SWAP gate with the control on the path
of the twin photon and SWAP on the polarization of it. Since we do not
choose any basis for measurement, the coincidence rate of the two detectors
is proportional to

\begin{equation}
Tr|\psi^{\prime}\rangle\langle\psi^{\prime}|\propto1+|\langle\psi_{a}|\psi
_{b}\rangle|^{2}\cos\varphi\text{.}
\end{equation}
In this way, we derive the visibility of the interferometer 
\begin{equation}
v=|\langle\psi_{a}|\psi_{b}\rangle|^{2}=Tr\rho_{a}\rho_{b}\text{.}
\end{equation}

We obtain this result based on two pure states $|\psi_{a}\rangle$ and $%
|\psi_{b}\rangle$, however, it can also be demonstrated that for input of
single-qubit mixed states $\rho_{a}$ and $\rho_{b}$, we still have the
relation $v=Tr\rho_{a}\rho_{b}$.

There are many possible ways of utilizing this result. For general
single-qubit states,\ we can measure the fidelity $\langle\psi|\rho
|\psi\rangle$\ of $\rho$\ with a pure state $|\psi\rangle$, overlap $%
Tr\rho_{a}\rho_{b}$\ between $\rho_{a}$ and $\rho_{b}$,\ purity\ $Tr\rho^{2}$
of a mixed state $\rho$ and Hilbert-Schmidt distance $d^{2}(\rho_{a},\rho
_{b})=1/2Tr(\rho_{a}-\rho_{b})^{2}$ between $\rho_{a}$ and $\rho_{b}$ \cite%
{E,F}.

Our experimental set-up is shown in Fig. 2(b). In the experiment, for
simplicity, we adjust HWP and tiltable\ QWP in the pump path to choose the
real numbers $a$ and $b$ in state (1). We use HWP and quartz in the signal
(idle) path to prepare input states. Two lenses ($f=300$ $mm$) are properly
placed to increase the collection efficiency of the single-mode fiber \cite%
{wang}.\ We add four QWPs in the set-up to compensate the\ phase shift
between horizontal and vertical polarization\ due to reflection. We use a
piezoelectric ceramics (PZT) in reflective mirror M$_{1}$ as a phase shift.
In the experiment, we first find the two HOM interferometers dip using the
coincidence of D$_{1}$ and D$_{4}$, D$_{2}$ and D$_{3}$. Then we use the
coincidence of D$_{2}$ and D$_{4}$ to observe the interference curve as we
adjust the voltage of PZT.

First, we input states into the network\ with the form $\rho_{a}\otimes
\rho_{b}$. In Fig. 3(a),\ We choose $\rho_{a}=|H\rangle\langle H|$\ and $%
\rho_{b}=(\cos2\theta|H\rangle+\sin2\theta|V\rangle)(\cos2\theta\langle
H|+\sin2\theta\langle V|)$ ($\theta$ is the angle between the optical axis
of the HWP and the vertical axis).\ In Fig. 3(b), we choose a pure$\ $state$%
\ \rho_{a}=(\cos2\theta|H\rangle+\sin2\theta|V\rangle)(\cos 2\theta\langle
H|+\sin2\theta\langle V|)$ and\ a mixed state$\ \rho _{b}=\left( 
\begin{array}{cc}
0.5 & 0.29 \\ 
0.29 & 0.5%
\end{array}
\right) $ which is prepared through a\ HWP ($\theta=22.5^{\circ}$)\ and a
6.5 $mm$\ quartz. From the figures, we can see that experimental dots are in
agreement with the theoretical curves. A similar experiment for measurement
of the overlap of two single qubit states can be seen in Ref. \cite{pla}.

For a general two-qubit state $\rho_{ab}$, we can obtain the relation 
\begin{equation}
v=Tr\rho_{ab}W=Tr\rho_{ab}^{T_{a}}|\Phi^{+}\rangle\langle\Phi^{+}|=\langle
\Phi^{+}|\rho_{ab}^{T_{a}}|\Phi^{+}\rangle\text{,}
\end{equation}
where $W$ is the SWAP operator in the form $|\Phi^{+}\rangle\langle\Phi
^{+}|^{T_{a}}$ ($|\Phi^{+}\rangle=$ $1/\sqrt{2}(|HH\rangle+|VV\rangle)$) and 
$T_{a}$ represents partially transposition about the system A. This result
agrees well with the Ref. \cite{F}. Then according to Peres-Horodecki
criterion \cite{ph}, if $v<0$, we can make sure that the state $\rho_{ab}$
is entangled. Hence the operator $W$ is just the entanglement witness
operator for these states. However, if $v>0$, we can't determine whether $%
\rho_{ab}$\ is entangled or not. In this way, we associate
entanglement-witness measurement with the sign of $v$. In addition, since
local unitary transformations cannot alter the entanglement of one state,
for all pure states and some kinds of mixed states, we can quantitatively
measure the entanglement of them with some proper local unitary
transformations. Moreover, for some special states, such as, the two-qubit
Werner state and the nonmaximally entangled state $a|HV\rangle\pm
b|VH\rangle $,\ the absolute values of $v$ are just equal to their
concurrence \cite{con}. In this sense, this scheme can be used for
quantitative entanglement-witness. So the set-up becomes an
\textquotedblleft entanglementmeter\textquotedblright\ for some partially
known states and may have important uses especially when the resources are
scarce.

In Fig. 3(c), the entangled input states are selected as $\cos 2\theta
|HH\rangle \pm \sin 2\theta |VV\rangle $. And as we expected, the visibility
keeps constant when we vary the value of$\ \theta $. However, In Fig. 3(d),
the input states become $\cos 2\theta |HV\rangle \pm \sin 2\theta |VH\rangle 
$, we obtain a sinusoidal curve which is in accordance with the value of
concurrence after experimental correction \cite{fig}.

Fig. 4 shows the entanglement-witness measurement. In order to demonstrate
the inverse of the interference curve of entangled state compared with that
of the disentangled state, in principle, we need to scan the whole
interference curves of the two cases. To overcome the difficulty of keeping
the phase stable for a long time, we first adjust the voltage of PZT and let
the phase stabilize at some place, such as $0$ or $\pi $. Next we replace
the entangled state with a disentangled one ($|H\rangle \otimes |H\rangle $)
to observe its interference. In order to ensure the phase stability during
the whole process, finally we return to the original entangled state. For
each input state, we consecutively note down 50 dots.\ For the three input
states in each figure, totally\ we obtain 150 dots.\ In Fig. 4(a) and (b),
we input\ the entangled state $\frac{1}{\sqrt{2}}(|HV\rangle -|VH\rangle )$
and\ we can see the obvious flip of the coincidence when replacing the input
states; while for Fig. 4(c) and (d), the input state is $\frac{1}{\sqrt{2}}%
(|HV\rangle +|VH\rangle )$, there are no obvious changes when input states
is alternated. So entanglement-witness measurement failed for this input
state.\ Though, we only demonstrate entanglement-witness measurement of the\
singlet state, for the other Bell states, we only need unilateral local
unitary transformations. We can generalize our set-up to\ other entangled
states such as, the Werner states. It just needs to prepare and input the
states before the Hong-Ou-Mandel interferometers.

As we know, for the general entanglement-witness measurement, first\ the
witness operators are decomposed into a pseudomixture of local operators,
then by adding the resulting expectation values of local operators with the
corresponding positive or negative weight, the value of witness operator is
obtained \cite{Guhne}. There also exists \textquotedblleft
structural\textquotedblright\ physical approximations (SPA) for estimating
the minimum eigenvalue of a transformed density matrix to detect quantum
entanglement according to positive maps separability criterion \cite{1,sss}
(also see \cite{buchong}).\ Compared with the method above and Ref \cite{F},
we find a method for direct detection of witness operators utilizing our
scheme and physically implement the positive but not completely positive
map, transposition, from another point\ of view.

In conclusion, we put forward a scheme to simulate the quantum network shown
in Fig. 1 using linear optical elements,\ which is based on two\
Hong-Ou-Mandel interferometers. As a basic building block, our scheme can be
cascaded to apply to high dimensional systems\cite{2003}.\ It can be used to
estimate functionals of high dimensional density matrices and avoid the
necessity of a complete state reconstruction. For example, it can evaluate
the spectrum, the extremal eigenvalues and eigenvectors of density matrices.
Based on this, the network may have potential applications in estimating
some properties of quantum channels, because a quantum channel is a trace
preserving linear map, which maps density operators into density operators 
\cite{E}. Recently, H. A. Carteret give a method that can determine the
spectrum of a partially transposed density matrix using the similar network
without the addition of noise to make the spectrum non-negative \cite%
{Carteret}.

Experimentally, we realize the scheme using linear optical elements. We
perform the measurement of overlap of two pure states and that of one pure
state and one mixed state. The results agree well with the theory. Our
set-up can be used to measure entanglement for all pure states and some
kinds of mixed states and implement entanglement-witness measurement for
states with non-positive partial transposition. We associate the value of
the interference visibility of the state with its concurrence to measure
entanglement and associate the inverse of the interference curve compared
with the one of the disentangled state with its entanglement to demonstrate
entanglement-witness measurement. We experimentally and directly detect the
value of witness operator and observe the obvious flip of the coincidence to
show their entanglement. From the other perspective, we physically implement
a positive but not completely positive map, transposition. However, usually
the map is indirectly implemented through SPA \cite{sss}. Using our set-up,
we realize the Fredkin (controlled-SWAP) gate, which may be helpful for
quantum information processing in the future.

The authors would like to thank Qin Wang, Bi-Heng Liu and Guo-Yong Xiang for
helpful discussions. This work was funded by the National Fundamental
Research Program (2001CB309300), National Natural Science Foundation of
China, Innovation Funds from Chinese Academy of Sciences, and Program for
New Century Excellent Talents in University.


\begin{thebibliography}{99}
\bibitem{fig} In Fig. 3(c), we derive the horizontal line according to the
experimental data, which should be $1$ according to theoretical calculation.
The main causes of the decrease of visibilities are mismatch of spatial
modes and non-perfect phase compensation. In Fig. 3(d), we incorporate the
experimental correction based on Fig. 3(c) into the theoretical curve.
\end{thebibliography}
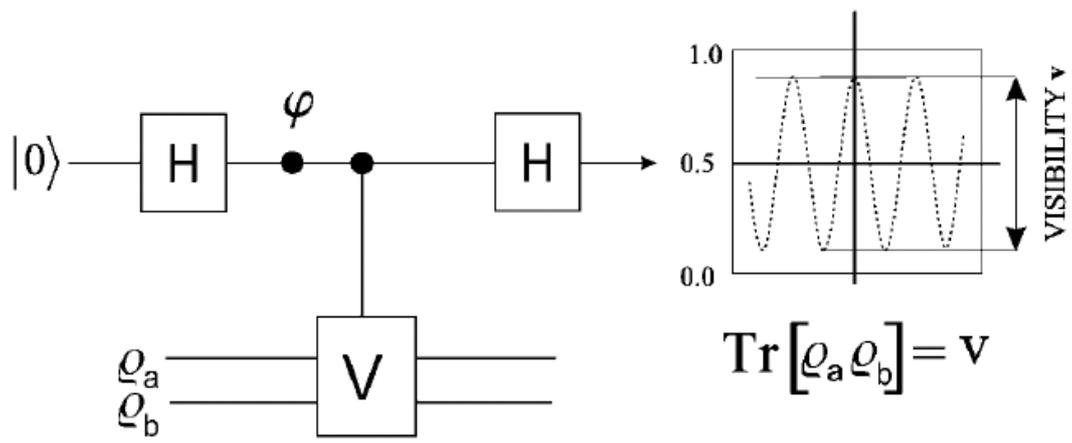
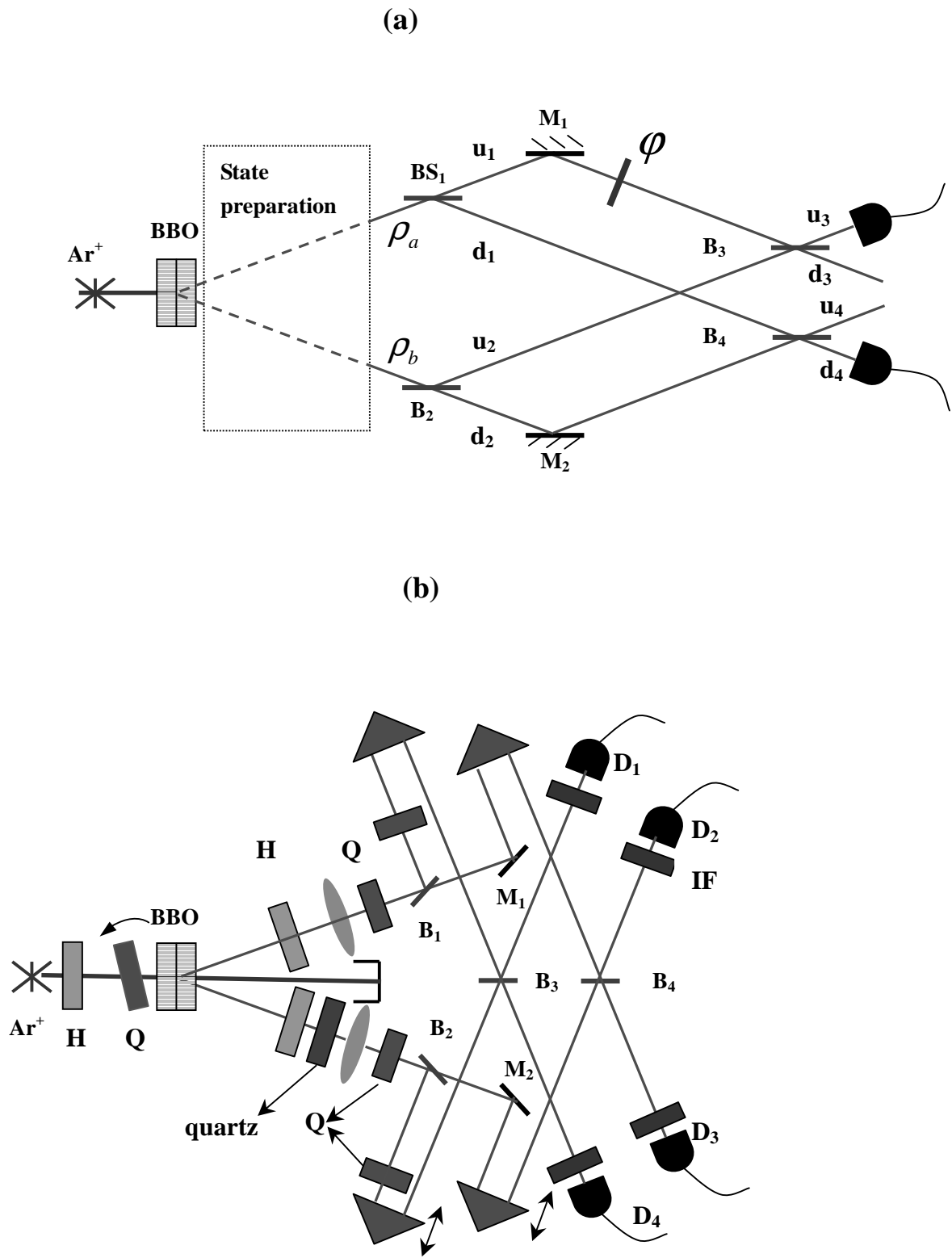
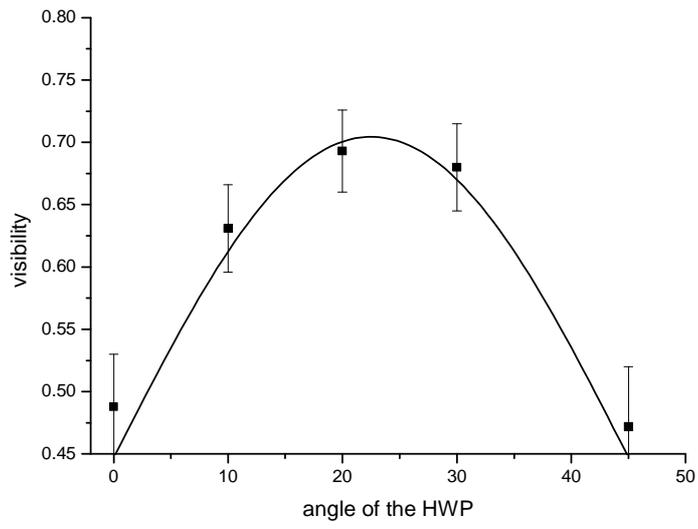
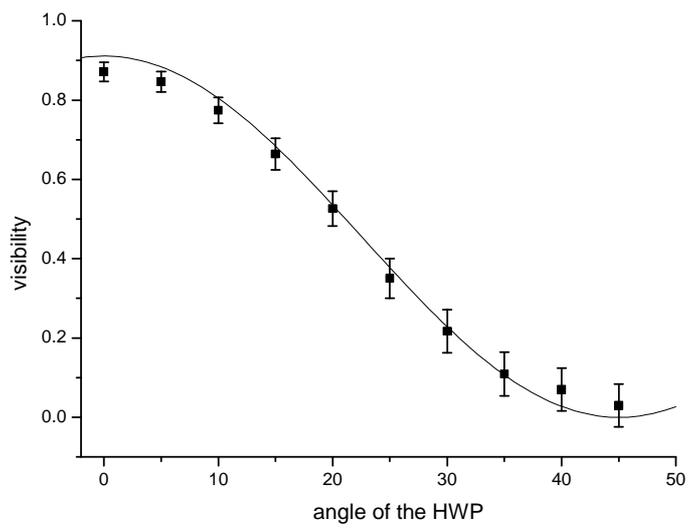
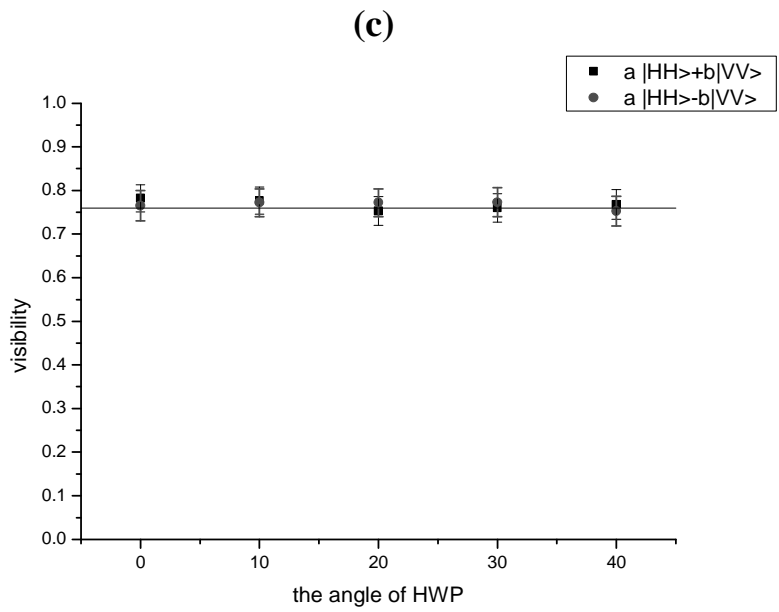
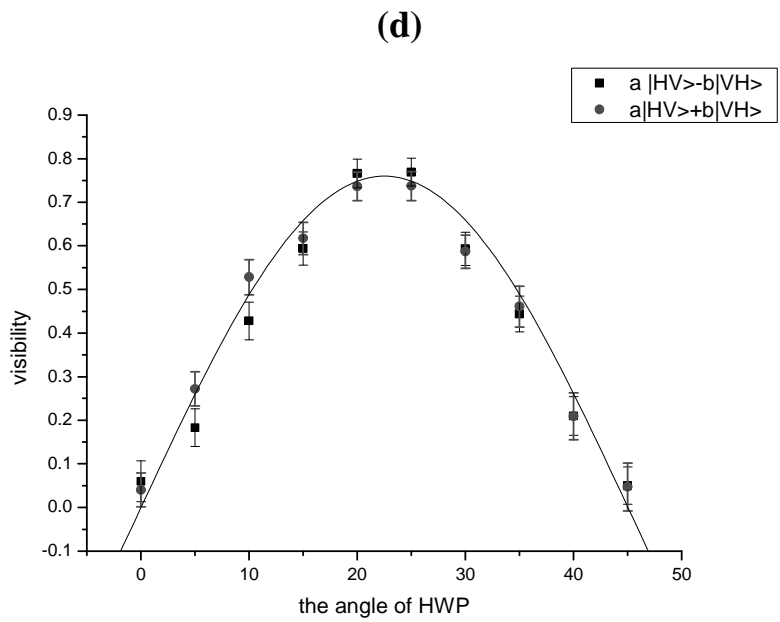
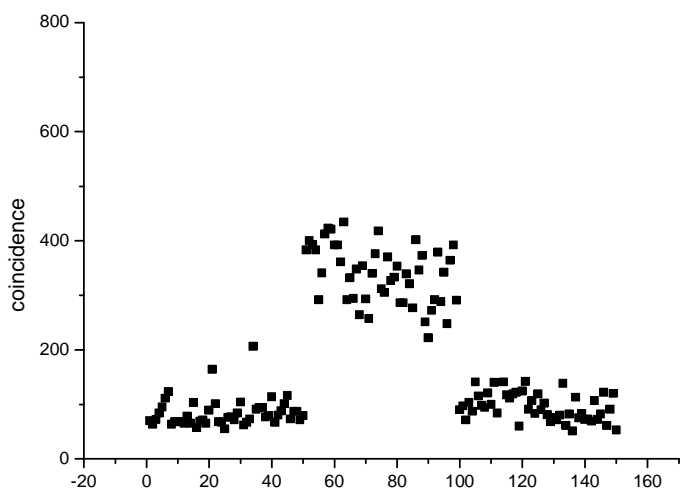
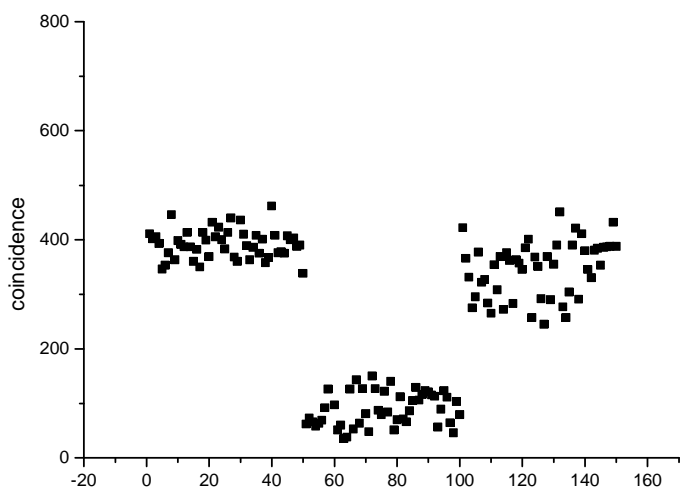
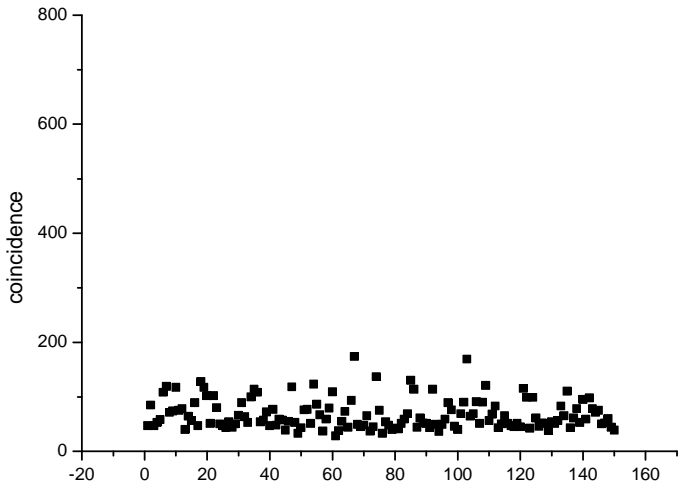
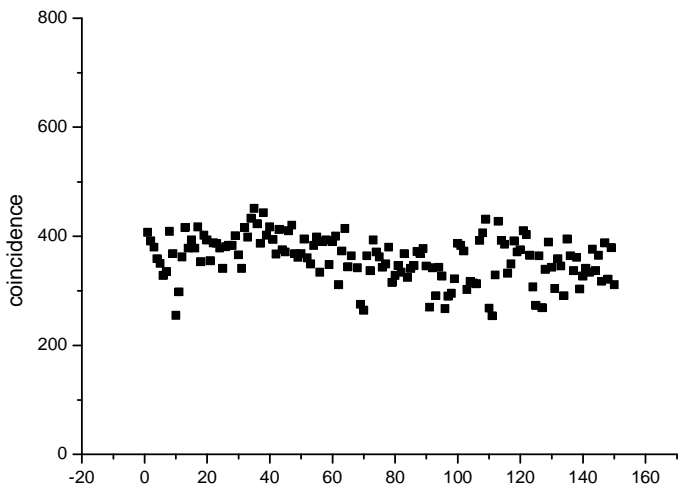
\end{document}